\documentclass[twocolumn,twoside]{revtex4}
\usepackage[utf8]{inputenc}
\usepackage{amsmath}
\usepackage{amsfonts}
\usepackage{amssymb}
\usepackage{graphics}
\usepackage{graphicx}
\usepackage{subfig}
\usepackage{pdfpages}
\usepackage{hyperref}
\usepackage{fancyhdr}
\begin{document}

\title{Measurement of CKM angle $\boldsymbol{ \phi_{3}} $ at Belle II}


\author{Niharika Rout\\
(On behalf of Belle II Collaboration)}
\affiliation{ Indian Institute of Technology Madras, Chennai, India, 600036}

\begin{abstract}
The CKM angle $ \phi_{3} $ is the only angle of the unitarity triangle that is accessible with tree-level decays in a theoretically clean way. The Belle II experiment is a substantial upgrade of the Belle detector and will operate at the SuperKEKB energy-asymmetric $ e^{+}e^{-} $ collider. The accelerator has already successfully completed the first phase of commissioning, with the first $ e^{+}e^{-} $ collisions recorded in 2018. The design luminosity of SuperKEKB is 8$ \times 10^{35}$ cm$^{-2}$s$^{-1}$ and the Belle II experiment aims to record 50 ab$ ^{-1} $ of data, a factor of 50 more than the Belle experiment. The key method to measure $ \phi_{3} $ is through the interference between $ B^{+} \to D^{0}K^{+} $ and $ B^{+} \to \overline{D}^{0}K^{+} $ decays, which occurs if the final state of the charm-meson decay is accessible to both the $ D^{0} $ and $ \overline{D}^{0} $ mesons. To achieve the best sensitivity, a large variety of $ D $ and $ B $ decay modes are required, which is possible at the Belle II experiment as almost any final state can be reconstructed, including those with photons. With the ultimate Belle II data sample of 50 ab$ ^{-1} $, a determination of $ \phi_{3} $ with a precision of 1$^{\rm o} $ or better is foreseen.
\end{abstract}

\maketitle

\thispagestyle{fancy}


\section{Introduction}
\begin{figure}\centering
\includegraphics[scale = 0.35]{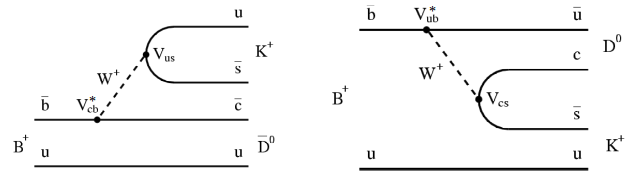}
\caption{Leading order quark flow diagrams for the decay channel $B^{+} \to DK^{+}$.}
\label{bdecay}
\end{figure}

The more precise determination of the $ CP $-violating  parameter $ \phi_{3} $ (also called $ \gamma $) is the most promising path to a better understanding of the Standard Model (SM) description of $ CP $ violation and search for contributions from non-standard model physics. It can be extracted via tree-level decays, along with non-perturbative strong interaction parameters, which makes the method free of theoretical uncertainties to $\mathcal{O}\left(10^{-7}\right)$~\cite{error}. Figure~\ref{bdecay} shows the two interfering diagrams for the most commonly used decay channel $ B^{\pm} \rightarrow DK^{\pm} $, where $ D $ indicates a $ D^{0} $ or $ \overline{D^{0}} $ meson decaying to the same final state $ f $; the weak phase $ \phi_{3} $ appears in the interference between $ b \rightarrow c\overline{u}s $ and $ b \rightarrow u\overline{c}s $ transitions. The $b \to u\bar{c}s$ amplitude ($A_{\rm sup}$) is suppressed relative to the $ b \to u\bar{c}s $ amplitude ($ A_{\rm fav} $) because of the magnitudes of the CKM matrix elements involved and the requirements of colorless hadrons in the final state. The two amplitudes  are related by
\begin{equation}
 \dfrac{A_{\rm sup}}{A_{\rm fav}} = r_{B}e^{i(\delta_{B} - \phi_{3})},
\end{equation}
where, $ r_{B} $ is the magnitude of the ratio of amplitudes and $ \delta_{B} $ is the strong-phase difference between the favoured and suppressed amplitudes. The current world average value of $ r_{B}$ is $ 0.103 \pm 0.005 $~\cite{hflav}.

\section{Primary methods to extract $\boldsymbol{\phi_{3}}$}
\subsection{GLW method}
In 1991, Gronau, London and Wyler (GLW) were the first to propose a method for measuring $ \phi_{3} $ using $ B^{\pm} \rightarrow DK^{\pm} $ decay, where the $ D $ decays to a $ CP $ eigenstate with eigenvalue $ \pm1 $~\cite{glw}. For the extraction of $\phi_{3}$, the following observables are used in the GLW method
\begin{equation}
R_{CP^{\pm}}  = 1 + r_{B}^{2} \pm 2 r_{B} \cos(\delta_{B}) \cos(\phi_{3}),
\end{equation}
\begin{equation}
A_{CP^{\pm}} = \pm 2 r_{B} \sin(\delta_{B}) \sin(\phi_{3})/ R_{CP^{\pm}}.
\end{equation}
$ CP $ eigenstates such as $ D\rightarrow K^{+}K^{-}$, $\pi^{+}\pi^{-}$ and $ K_{\rm S}\pi^{0} $  are used to extract $ \phi_{3} $ via this method.
\subsection{ADS method}
This method was proposed by Atwood, Dunietz and Soni (ADS) in 1997~\cite{ads}. The main idea was to pick a final state for which $D^{0}\rightarrow \textit{f}$ is suppressed relative to $\overline{D^{0}}\rightarrow \textit{f}$. For example, $B^{-}\rightarrow[K^{+}\pi^{-}]K^{-}$ can be reached via doubly Cabibbo-suppressed decay mode $D^{0}\rightarrow K^{+}\pi^{-}$ or via Cabibbo-favored decay mode $\overline{D^{0}}\rightarrow K^{+}\pi^{-}$. The observables used are
\begin{equation}
R_{\textit{ADS}} = r_{B}^{2} + r_{D}^{2} + 2r_{B}r_{D} \cos(\delta_{B}+\delta_{D})\cos(\phi_{3}),
\end{equation}
\begin{equation}
A_{\textit{ADS}} = 2r_{B}r_{D}\sin(\delta_{B}+\delta_{D})\sin(\phi_{3})/R_{\textit{ADS}}.
\end{equation}
Here, $r_{D}$ and $\delta_{D}$ are the amplitude ratio of the suppressed and favored $ D $ decays and the strong-phase difference between them, respectively.
\subsection{GGSZ method}
\begin{figure}[h!]
\centering
\includegraphics[scale = 0.45]{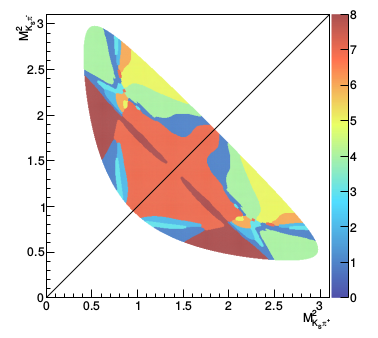}
\caption{Optimal binning of $ D \to K_{\rm S}^{0}\pi\pi $ Dalitz plot.}
\label{dalitz}
\end{figure}

This method was proposed by Giri, Grossman, Soffer and Zupan in 2003~\cite{ggsz}. The method uses self-conjugate multi-body $ D $ final states, such as $ K_{\rm S}^{0}\pi^{+}\pi^{-}$ and $ K_{\rm S}^{0}K^{+}K^{-} $. In this method, the $ D $ Dalitz space is binned in a way that gives the maximum sensitivity to $ \phi_{3} $ in a model-independent manner. The binning eliminates the model-dependent systematic uncertainty in the measurement and can give degree-level precision~\cite{B2TIP}. Figure~\ref{dalitz} shows an optimal binning used for a GGSZ analysis~\cite{B2TIP}. The signal yield in each bin is given by

\begin{equation}
\Gamma_{i}^{\pm} \propto K_{i} + r_{B}^{2}\overline{K}_{i} + 2\sqrt{K_{i}\overline{K}_{i}}(c_{i}x_{\pm} + s_{i}y_{\pm}),
\end{equation}
where $(x_{\pm}, y_{\pm}) = r_{B}(\cos(\pm \phi_{3} + \delta_{B}), \sin(\pm \phi_{3}+ \delta_{B}))$. Here, $ K_{i} $ is the number of events in the $ i^{\rm th} $  bin of a flavour tagged $ D $ decay sample; these parameters are obtained with high precision using a large statistics sample of $ D^{*\pm} \rightarrow D\pi^{\pm} $ decays. The parameters $ c_{i} $ and $ s_{i} $ are the amplitude-averaged strong phase difference between $ \overline{D^{0}} $ and $ D^{0} $ over $ i^{\rm th} $ bin and can be measured using quantum correlated pairs of $ D $ mesons created at $ e^{+}e^{-} $ annihilation experiments operating at the threshold of $ D\overline{D} $ pair production~\cite{libby}. The $(x_{\pm}, y_{\pm})$ parameters can be obtained from equation (6) using maximum likelihood method.

The average value of $ \phi_{3} $ obtained when combining all measurements from the Belle collaboration is $(73^{+13}_{-15})^{\rm o}$, which is dominated by the GGSZ final states. The current world average value of $ \phi_{3} $ is $(73.5^{+4.2}_{-5.1})^{\rm o}$, where the precision is dominated by the results from LHCb experiment.

\section{SuperKEKB and Belle II detector}

The SuperKEKB colliding-beam accelerator provides $e^{+}e^{-}$ collisions at an energy corresponding to the mass of the $\Upsilon(4S)$ resonance, which are being recorded by the Belle II detector. SuperKEKB consists of two storage rings of 3.012 km length each, one for the 7 GeV electrons and one for the 4 GeV positrons. The design peak instantaneous luminosity of SuperKEKB is 8 $\times$ 10$^{35}$ cm$^{-2}$s$^{-1}$, approximately forty times higher than what was achieved by the KEKB accelerator. This will allow a data sample to be accumulated that corresponds to an integrated luminosity of  50~ab$^{-1}$.

\par
\begin{figure}[h!]
\centering
\includegraphics[scale = 0.25]{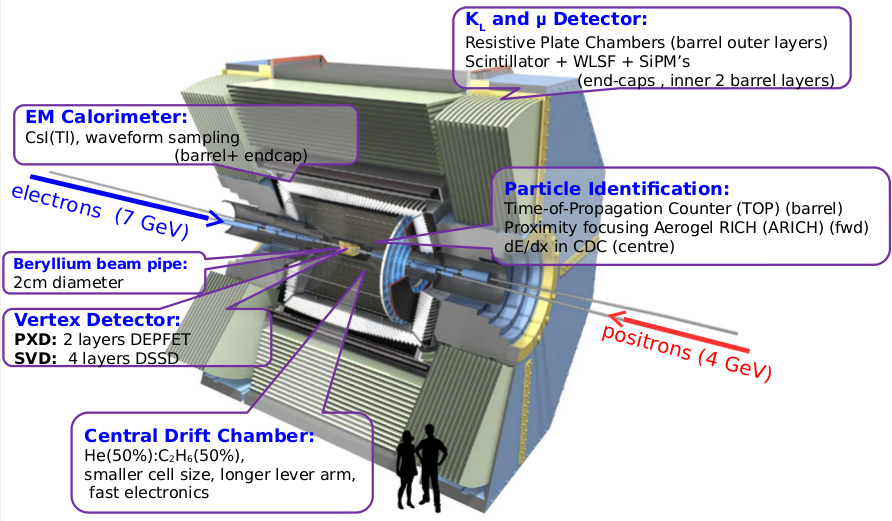}
\caption{Belle II detector.}
\label{detector}
\end{figure}

 Belle II is the upgraded version of the Belle detector. It can tolerate the much higher level of beam-related background that arises from the increase in instantaneous luminosity. The different subdetectors are shown in the Fig.~\ref{detector}. In terms of performance, it has good tracking, vertexing, $ K $-$ \pi $ separation and good neutral reconstruction efficiency.

\section{Glimpse of the phase II data collected by Belle II}
The accelerator commissioning of the Belle II experiment, also known as Phase I, was completed in 2016. The detector entered its second commissioning period (Phase II) in February 2018, with the first collisions taking place on the 25$ ^{\rm th} $ April, 2018. A data sample was collected corresponding to an integrated luminosity of 0.5 fb$ ^{-1} $. Only one ladder of each layer of the vertex detector was present during the data taking, which corresponds to $\frac{1}{8} $ of the full detector.
\par
\begin{figure}[!ht]
\begin{center}
\includegraphics[width=8cm]{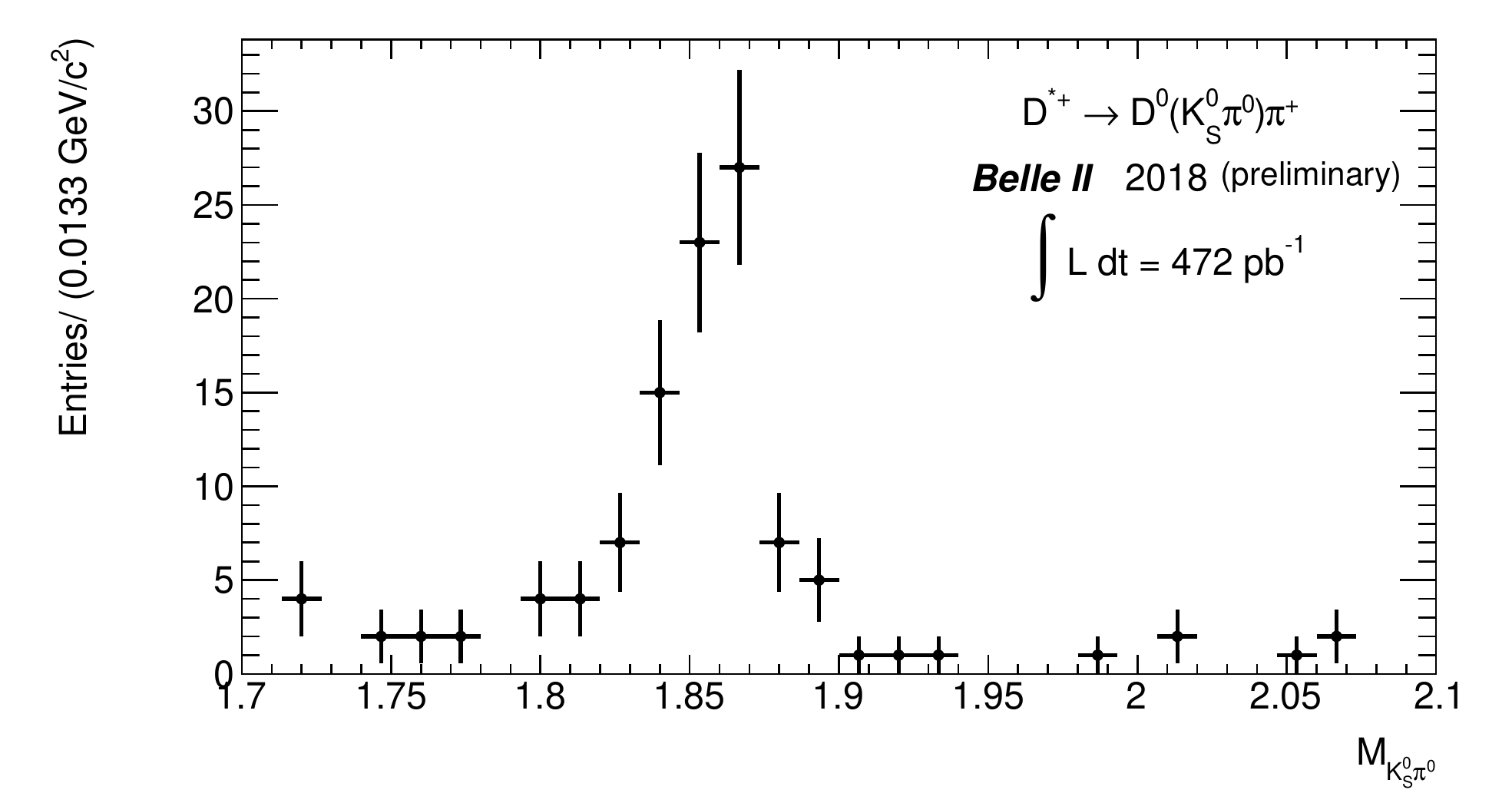} 
\caption{$M(K_{\rm S}^{0}\pi^{0})$ for the mode $ D^{*\pm} \rightarrow D^{0} (K_{\rm S}^{0}\pi^{0})\pi^{\pm} $.}
\label{fig:kspi0}
\end{center}
\end{figure}

\begin{figure}[!ht]
\begin{center}
\includegraphics[width=8cm]{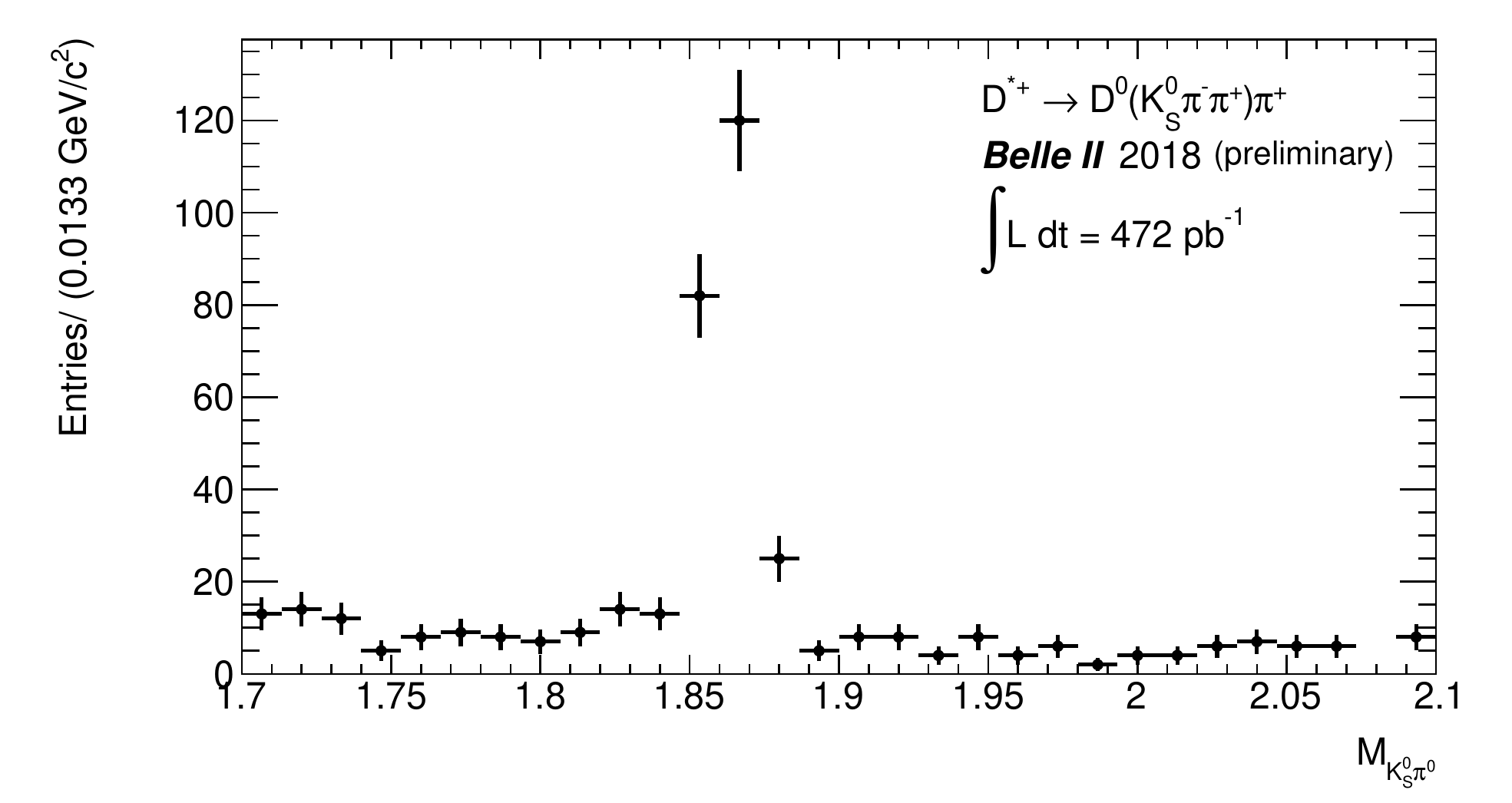} 
\caption{$M(K_{\rm S}^{0}\pi^{-}\pi^{+})$ for the mode $ D^{*\pm} \rightarrow D^{0} (K_{\rm S}^{0}\pi^{-}\pi^{+})\pi^{\pm} $.}
\label{fig:kspipi}
\end{center}
\end{figure}

\begin{figure}[!ht]
\begin{center}
\includegraphics[width=7cm]{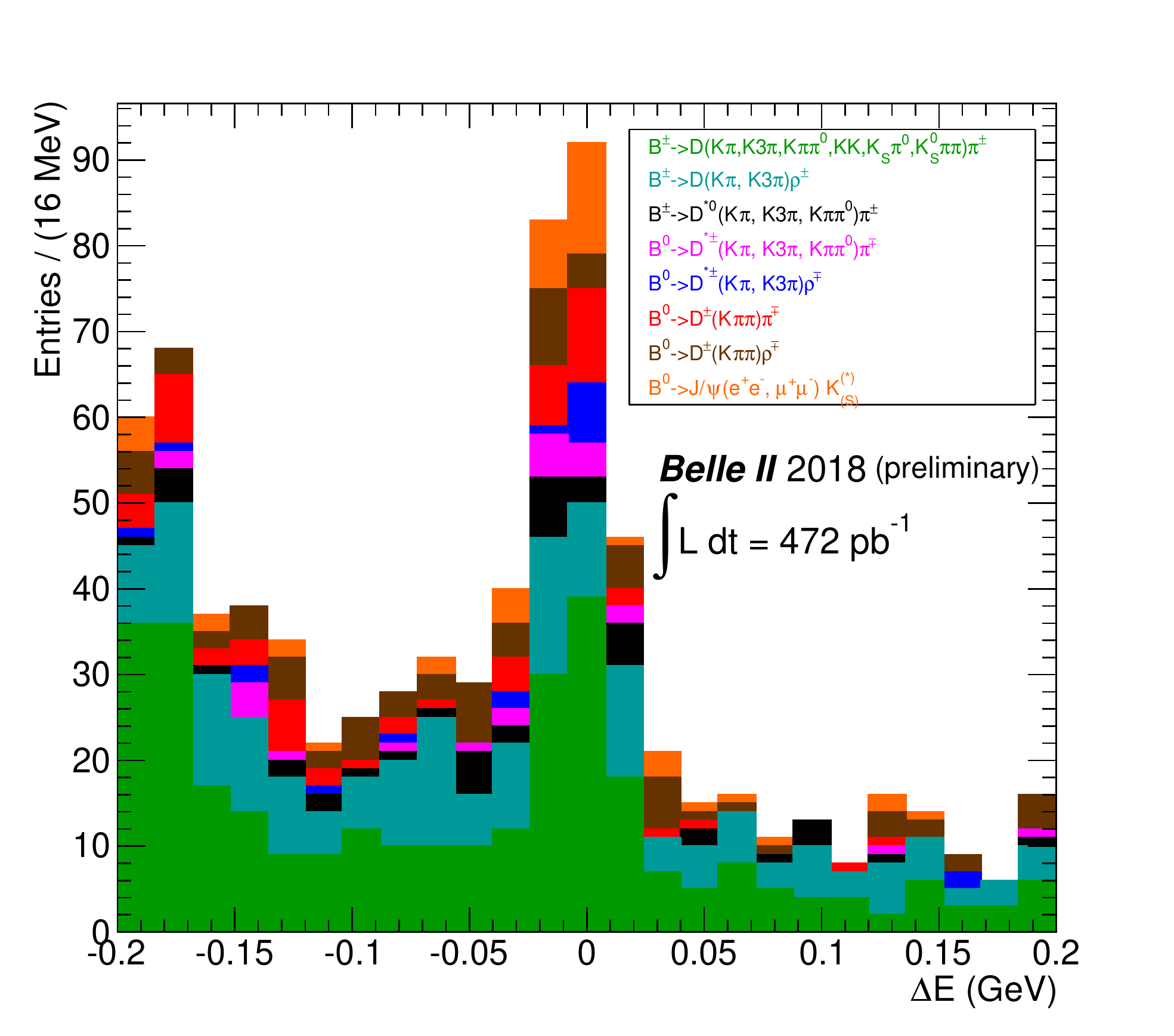} 
\caption{ $\Delta E$ distribution for the $B$ candidates selected in the Phase II data.}
\label{fig:btodpi}
\end{center}
\end{figure}

 We observed various charm decay modes, including the $ CP $ eiganestates, like $ K_{\rm S}^{0}\pi^{0} $ and multi-body final states like $ K_{\rm S}^{0}\pi^{+}\pi^{-} $, validating the potential for charm physics at Belle II. The invariant mass distributions for $ D^{0} \to K_{\rm S}^{0}\pi^{0}$ and $ D^{0} \to K_{\rm S}^{0}\pi^{-}\pi^{+} $ are shown in Figs.~\ref{fig:kspi0} and~\ref{fig:kspipi}, respectively. Possible charged and neutral $ B $ meson candidates are reconstructed from different charmed mesons and a total of 245 signal events are obtained, which is consistent with the yields from the data samples of the ARGUS/CLEO experiment \cite{argus}. The signal enhanced $\Delta E$ distribution is shown in Fig.~\ref{fig:btodpi}, which is defined as
\begin{equation}
\Delta E = \Sigma E_{i} - E_{\rm beam} ,
\end{equation}
where $ E_{\rm beam} $ is the beam energy in the center-of-mass frame and $ E_{i} $ is the energy of the $ B $ daughter particles in the center-of-mass frame. 

\vspace{-0.3in}

\section{$\boldsymbol{ B \to DK }$ decays at Belle II}

The current level of precision of the measurements of $ \phi_{3} $ is dominated by the statistical uncertainty from the limited number of $ B $ decays. The 50~ab$ ^{-1} $ data set of Belle II will improve the uncertainty significantly. The first $ \phi_{3} $ sensitivity study at Belle II was performed using the mode $ B^{+} \to D (K_{S}^{0}\pi^{-}\pi^{+}) K^{+} $ with the GGSZ formalism. Based on these studies a 3$ ^{\rm o} $ precision is expected from the 50~ab$ ^{-1} $ data set using this mode alone~\cite{B2TIP}. However, the anticipated precision is 1.6$ ^{\rm o} $ when all Belle results, including GLW and ADS as well, are extrapolated to the 50~ab$ ^{-1} $ data-set.

\par
\begin{figure}[!ht]
\begin{center}
\includegraphics[width=7cm]{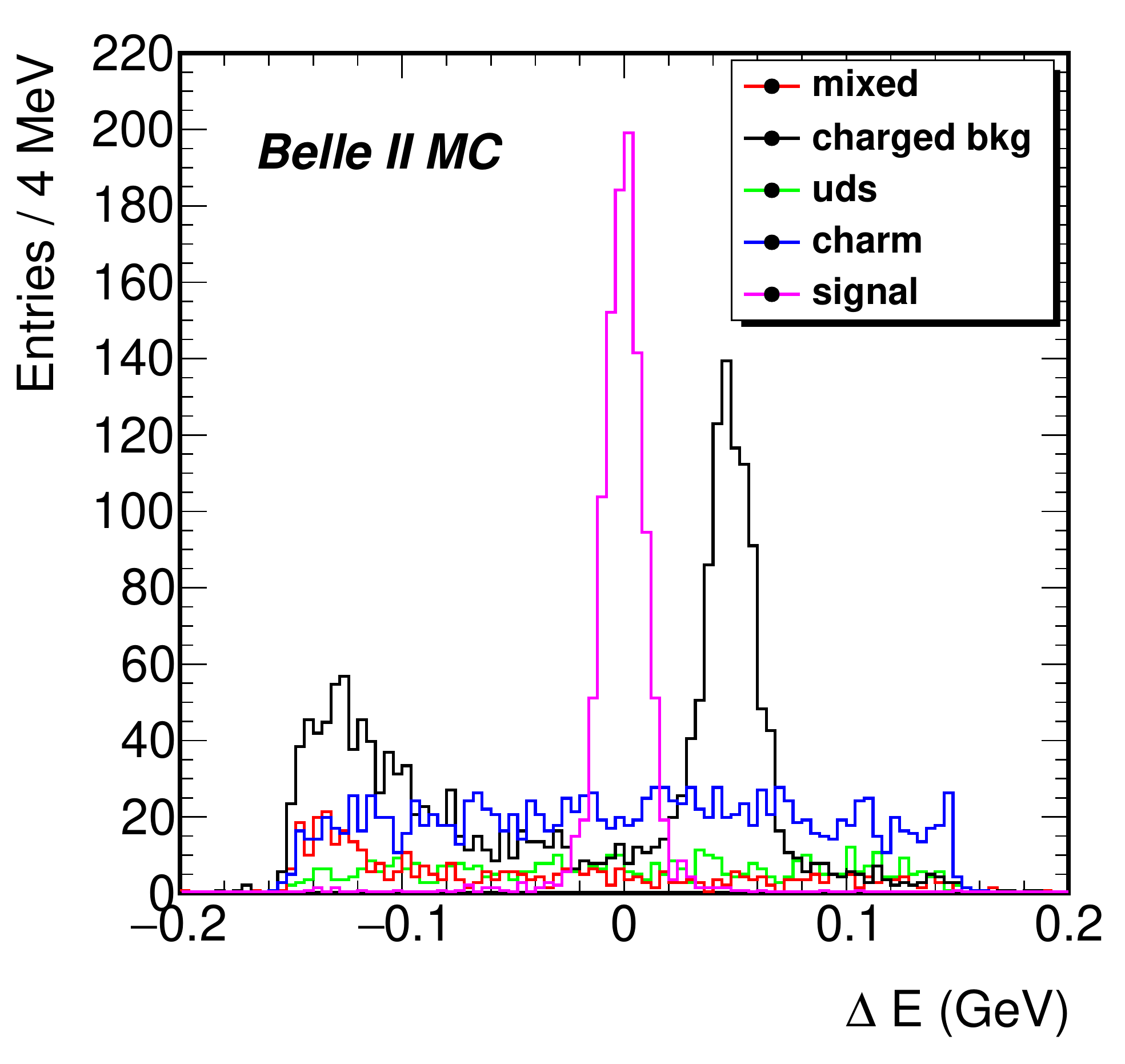} 
\caption{ $\Delta E$ distribution using MC after continuum suppression at Belle II. The peak at 40~MeV is from the topologically identical $ B^{+} \to D\pi^{+} $ events in which a $ \pi $ has been misidentified as kaon.}
\label{fig:cs}
\end{center}
\end{figure}

 At Belle II, the dominant background comes from $ e^{+}e^{-} \rightarrow q\overline{q} $, where $ q = u, d, s, c $, also known as continuum events. So, continuum suppression plays an important role in identifying the signal from the huge background. Figure~\ref{fig:cs} demonstrates the outcome of continuum suppression algorithm at Belle II for the mode $ B^{+} \to D (K_{S}^{0}\pi^{-}\pi^{+}) K^{+} $ using Belle II Monte Carlo (MC) sample.

We are currently analyzing the following decay modes at Belle II.

\begin{itemize}
\item $ B^{+} \to D (K_{S}^{0}\pi^{-}\pi^{+}) K^{+} $,
\item $ B^{+} \to D (K_{S}^{0}\pi^{0}) K^{+} $ and
\item $ B^{+} \to D (\pi^{-}\pi^{+},K^{+}K^{-}) K^{+} $.
\end{itemize}

The improved particle identification, good neutral and $ K_{S}^{0} $ reconstruction, better tracking efficiency and improved continuum suppression algorithms at Belle II will all benefit the selection of these modes. In addition to these modes, $ D^{0} $ hadronic parameters measured at external charm factories like \textit{BESIII} will play a vital role.

\section{Future prospects}
\begin{figure}[!ht]
\begin{center}
\includegraphics[width=8cm]{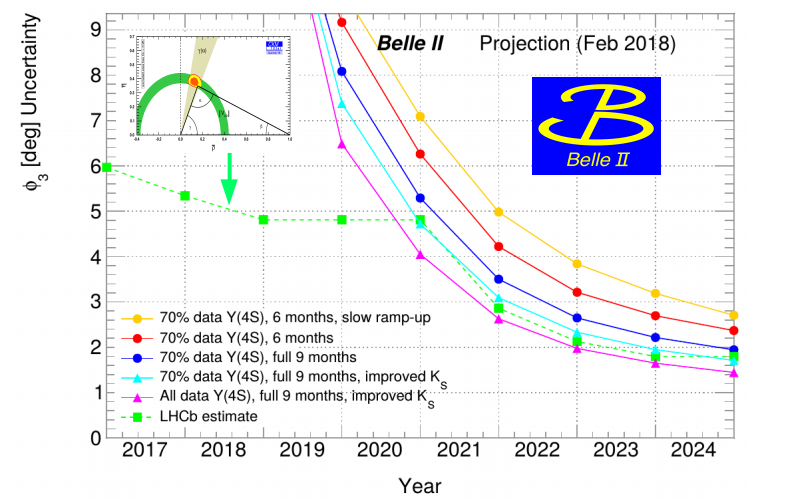} 
\caption{Belle II projection of expected uncertainty on $ \phi_{3} $ by 2027 as per the new luminosity projection of SuperKEKB~\cite{skekb}.}
\label{fig:proj}
\end{center}
\end{figure}

The phase III run of Belle II has already started. We expect Belle II and LHCb upgrade to match each others performance. Due to the unbiased trigger, Belle II will give excellent performance for Dalitz plot analyses. In addition, sensitivity to the neutral particles will allow the inclusion of more $ D $ modes, though LHCb will clearly have more precise results in final states consisting solely of charged tracks. Figure~\ref{fig:proj} shows the precision of $ \phi_{3} $ with the data sets that will be collected from 9 months run of Belle II experiment. So, we clearly expect the uncertainty to be less than 2$ ^{\rm o} $ by 2027.

\section{Summary}
\begin{figure}[!ht]
\begin{center}
\includegraphics[width=8.5cm]{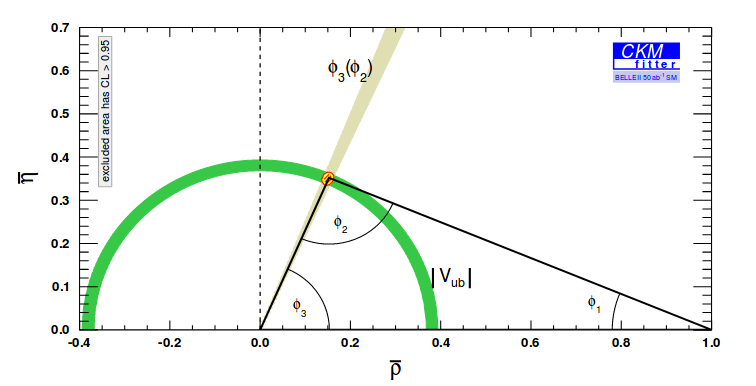}
\caption{Fit extrapolated to the 50 ab$ ^{-1} $ for an SM-like scenario.}
\label{fig:ckm}
\end{center}
\end{figure}

The precise measurement of the angle $ \phi_{3} $ will give us a SM benchmark to which other measurements of the CKM parameters can be referred to, both within the SM and beyond. The current uncertainty on $ \phi_{3} $ is $ \sim $ 5$ ^{\rm o} $. Combined sensitivity of 1.6$ ^{\rm o} $ is expected when all Belle results are extrapolated to 50~ab$ ^{-1} $ data-set. Accounting for recent results of new physics in tree-level amplitudes, a shift of up to 4$ ^{\rm o} $ on the SM value of $ \phi_{3} $ is possible~\cite{np}. This is one of the strongest motivations for the 1$ ^{\rm o} $ precision being pursued by Belle II. Figure~\ref{fig:ckm} shows the precision on $ \phi_{3} $ in CKM triangle when the fit extrapolated to 50~ab$ ^{-1} $ for a SM-like scenario~\cite{B2TIP}.
\par The phase II run of Belle II was successful despite of the very small data sample collected. Phase III run is in full swing and Belle II is ready to realize its potential for flavor physics.

\end{document}